\documentclass[preprint,12pt]{aastex}

\begin{document}

\title{The Formation of Population III Binaries}

\author{Kazuya Saigo\altaffilmark{1}, Tomoaki Matsumoto\altaffilmark{2}, \&  Masayuki Umemura\altaffilmark{3}}

\altaffiltext{1}{National Astronomical Observatory of Japan, Mitaka, Tokyo 181-8588, Japan; saigo@th.nao.ac.jp }

\altaffiltext{2}{Department of Humanity and Environment, Hosei University, Fujimi, Chiyoda-ku, Tokyo 102-8160, Japan; matsu@i.hosei.ac.jp }

\altaffiltext{3}{Center for Computational Sciences, University of
  Tsukuba, Japan; umemura@rccp.tsukuba.ac.jp}


\begin{abstract}
We explore the possibility for the formation of Population III binaries. 
The collapse of a rotating cylinder is simulated with a three-dimensional, 
high-resolution nested grid, assuming the thermal history of primordial gas. 
The simulations are done with dimensionless units, 
and the results are applicable to low-mass as well as massive
systems by scaling with the initial density. 
We find that if the initial angular momentum is as small as
$\beta \approx 0.1$, where $\beta$ is the ratio of centrifugal force 
to pressure force, then the runaway collapse of the cloud stops to
form a rotationally-supported disk. After the accretion of the envelope, 
the disk undergoes a ring instability, eventually fragmenting into a binary. 
If the initial angular momentum is relatively large, 
a bar-type instability arises, resulting in the collapse into
a single star through rapid angular momentum transfer. 
The present results show that
a significant fraction of Pop III stars are expected to form
in binary systems, even if they are quite massive or less massive.
The cosmological implications of Population III binaries are briefly
discussed.

\end{abstract}


\section{Introduction}
\label{intro}

In present-day galaxies, it is known that more than half of main-sequence 
stars are found in binaries or low-order multiple systems not only for low-mass
stars but also for massive stars (e.g., Tohline 2002; Kroupa 2004). 
Regarding the upper mass limit, a very massive binary of the order of 
$100M_\odot$ is found in the Large Magellanic Cloud \citep{Ostrov02}.
Furthermore, it has been revealed that the binary fraction of 
pre-main-sequence stars is comparable to that of main-sequence stars 
\citep{ Mathieu1994, Prosser94}. This implies that binaries possibly form prior to the pre-main-sequence phase.

Recent studies on the formation of Population III stars
have shown that the first stars in the universe are likely to 
be as massive as $100-1000M_\odot$ 
\citep{BCL99, Abel00, Abel02} or to
have formed in a bimodal initial mass function with peaks of $\sim 1M_\odot$ and 
several $100M_\odot$ \citep{NU99, NU01}. 
In light of present-day star formation, 
a portion of Population III stars appear able to form in binary systems. 
Population III binaries could play an important role for the thermal history
of the universe through the formation of massive binary black holes (BHs), 
accreting binaries, and Population III Type Ia supernovae (SNe).
For instance, the accreting binaries may emit ultraviolet radiation in
a longer timescale than the lifetime of massive stars 
themselves and therefore can be a reionization source in
the early universe \citep{Umemura04}.

The binary formation for present-day stars has 
been studied through elaborate numerical simulations
\citep{Bonnell94,Boss86,Tsuribe99,MH03,NL03,MTM04}.
However, no study has been made on the formation of
Population III binaries. So far, studies of Population III protostars have focused on
the core collapse of single stars which proceeds through the cooling by primordial
hydrogen molecules \citep{PSS83,ON98,Omukai00}. 
The previous studies imply that 
(1) the initial stellar core is on the order of $M_{\rm core}=10^{-2} M_\odot $,
which is not so different from the present-day protostars, 
(2) the mass accretion rate is as high as a few tens of $c_s^3/G$
with the sound speed $c_s$, which is numerically
$\dot{M}\approx 10^{-3} -10^{-1} M_\odot {\rm yr}^{-1}$
for $\approx 100$K gas because no coolant works to cool down to less than $100$K in primordial gas, and 
(3) the thermal history follows the evolution with an adiabatic exponent 
of $\gamma \approx 1.1$ until reaching the central density of 
$n \approx 10^{20}{\rm cm}^{-3}$, above which hydrogen molecules
are dissociated and therefore the temperature increases with $\gamma=5/3$.

On the basis of these findings, we study the formation of Population III binary stars.
We perform a three-dimensional hydrodynamic simulation 
with nested grids in order to resolve the core collapse and 
the subsequent fragmentation. 
In the next section, the numerical model is described. 
In \S \ref{Results}, we present the numerical results for 
the collapse and fragmentation of Population III stellar core. 
\S \ref{Discussion} is devoted to the discussion of 
the cosmological implications of Population III binaries.

\section{Model}
\label{numsim}

In the previous three-dimensional numerical simulations of the formation of Population III stars \citep[e.g.,][]{BCL99}, filamentary structure
commonly appears before it is broken into many pieces. 
Hence, we consider a cylindrical cloud as the initial condition, in which the cloud is assumed to be in hydrostatic balance with the central temperature and density of 
$T_{\rm c0}=800$K and $n_{\rm c0}=1\times 10^{12}{\rm cm}^{-3}$,
respectively. 
The equation of state is assumed to be polytropic with $\gamma = 1.1$. 
The polytropic constant $K$ is proportional to $n_{\rm c0}^{1 - \gamma} 
\, T_{\rm c0}$ and fixed to $ \, 6.87 \times 10^{7}$ in SI units throughout the calculation. Therefore, if a different initial density is adopted, the temperature
is scaled as $T_{\rm c0}=800{\rm K}(n_{\rm c0} /10^{12}{\rm cm}^{-3})^{0.1}$.
We add a 5\% sinusoidal density fluctuation of one critical wavelength along the cylinder axis, above which the cylinder is gravitationally unstable. 
In addition, in the azimuthal direction, a perturbation of $m=2$ and 4 mode
is blended with $\delta \rho/\rho=10\%$. 
The total mass for one critical wavelength is $M=1.5M_\odot$
for the assumed initial density.
We note that, since actual simulations are done with dimensionless units,
the present model follows the scaling as 
$M=1.5M_\odot (n_{\rm c0}/10^{12}{\rm cm}^{-3})^{(3\gamma -4)/2}$
for the system mass and $L=1477{\rm AU} (n_{\rm c0}/10^{12}{\rm cm}^{-3})^{(\gamma-2)/2}$ for the linear size of computational domain.
For instance, the numerical results can be translated 
into a massive case with $M=423M_\odot$ and $L=1.04{\rm pc}$,
if we renormalize the simulation by $n_{\rm c0}=1\times 10^{5}{\rm cm}^{-3}$. 

The rotation of the initial cloud is specified by a model parameter $\beta$, which is defined by the ratio of the centrifugal force to the pressure gradient force, i.e., 
$v_\phi^2 / r = - \beta  (1/\rho) (\partial P / \partial r)$.
For a given $\beta$, the specific angular momentum at the outer 
boundary, $R$, is given by $j=[2Gl R^2 \beta/(1+\beta)]^{1/2}$,
where $l$ is the mass per unit length. 
We construct four models with different rotation of $\beta \, = \,$ 0.0, 0.1, 0.3 and 1.0.

The hydrodynamic evolution of the cloud is solved by a three-dimensional nested grid
code, which is developed by \citet{MH03}.
At the initial epoch, the cylinder is represented in 
$(n_x, n_y, n_z) = (128, 128, 32)$ grid points,
assuming a mirror symmetry with respect to the equatorial plane.
We introduce a new finer grid so that the resolution satisfies the condition that the half-width of the half-maximum density should be resolved with more than 8 grid points.
Here 16 levels for nested grids are allowed, and therefore
the final spatial resolution 
$\Delta x = L /(128 \times 2^{15}) \simeq L/(4.2 \times 10^6)$.
Here we set $L=1477$AU and therefore $\Delta x =3.5 \times 10^{-4}$AU. 
The self-gravity is calculated self-consistently by solving the Poisson
equation with a multigrid iteration method \citep{MH03a}.

\section{Results}
\label{Results}

The numerical results can be divided into two cases of low and high $\beta$ cases. 
Here, we show the typical results of these models.

\subsection{Low $\beta$ case}

Figure 1 shows the density distribution and velocity fields for a case of $\beta =0.1$.
At the initial stage (Fig. 1a), the ratio of rotation energy
$E_{\rm rot}$ to gravitational energy $|W|$ is 
as small as $E_{\rm rot}/|W|=0.0187$. 
The filamentary cloud fragments as a results of gravitational instability, and each fragment undergoes the runaway collapse.
In early stages of runaway collapse, the dense cloud shrinks, keeping an approximately spherical shape. 
At the stage of $n_{\rm c} \, = \, 1.04 \times 10^{14}$ cm$^{-3}$, $E_{\rm rot}/|W|$ of spherical core reaches 0.245 because of the spin-up of the central cloud during the collapse.
After that, the spherical collapsing core forms a runaway collapsing disk in the central region (Fig. 2b). 
When the cloud collapses near to the centrifugal barrier, 
a rotationally supported disk, the radius of which is $\simeq 0.035$ AU is formed (Fig. 1c) at the stage of $n_{\rm c} \, = \, 1.82 \times 10^{19}$ cm$^{-3}$. 
The formation of a rotationally-supported disk 
is an important consequence for the binary formation.
If the equation of state is perfectly isothermal ($\gamma=1$), 
the runaway collapse of a rotating cloud 
continues to form the central singularity without 
forming a rotationally supported disk \citep{SH98,SHM00}.
However, the present simulation shows that the runaway collapse
stops before producing the central singularity in the case of $\gamma=1.1$. 
This is an essential difference between isothermal and $\gamma=1.1$ evolution. 
After a rotationally supported disk forms, the envelope accretes 
onto the disk (Fig. 1d). 
In this disk, the Toomre $Q (= \kappa c_s / \pi G \sigma)$
is slightly less than unity. 
However, the disk radius is smaller than the critical wavelength of ring instability.
The disk is therefore stable at this stage. 
The disk radius and mass grow via accretion from the infalling envelope. 

When the radius increases up to 0.1 AU, which is larger than the critical wavelength, the disk suffers from ring instability (Fig. 1e).
The rotating ring is unstable against nonaxisymmetric modes,
finally evolving into a binary core (Fig. 1f).
The binary core separation is $\approx 0.1$AU. 
The simulation was terminated just before the central density reaches 
$\approx 10^{20}{\rm cm}^{-3}$.

The subsequent equation of state is expected to deviate 
from $\gamma=1.1$ to $5/3$. 
When the binary core forms, the core mass is approximately $10^{-2}M_\odot$.
The binary core can continue to grow through further accretion.
The accretion rate onto the binary system is 
$\dot{M}\approx 1 \times 10^{-1}M_\odot {\rm yr}^{-1}$.
Saigo \& Hanawa (1998) derived a Larson-Penston type accretion rate 
onto a disk as $\dot{M}=8.15 c_s^3/G$.
The present accretion rate is comparable to this prediction
for the temperature of $\sim 5,000$K which is attained by 
the stage of $n \sim 1 \times 10^{17}$ cm$^{-3}$. 
Except for the formation of a binary core,
the mass accretion after core formation is likely to be similar to the case
of a single Population III star \citep{ON98}.

For another low $\beta$ case, we have simulated
a $\beta =0.3$ model as well. As a result, we have found that
the dynamical evolution is very similar to the case of $\beta =0.1$.
Eventually, a binary core forms.

\subsection{High $\beta$ case } 

Models with $\beta \, = \, 1.0$ exhibit 
the evolution different from the low $\beta$ models 
described in the previous section. 
Figure 2 shows the evolution of a high $\beta$ case. 
At the initial stage (Fig. 2a), $E_{\rm rot}/|W|=0.103$. 
The runaway collapse proceeds in a manner similar to the case
of $\beta =0.1$ (Fig. 2b).
Until a nearly spherical collapsing cloud forms, 
$E_{\rm rot}/|W|$ increases up to $E_{\rm rot}/|W|=0.327$,
owing to the enhanced rotation energy. Then 
a bar mode structure emerges as shown in Figure 2c,
because the bar mode instability can set in when $E_{\rm rot}/|W|>0.27$ 
\citep[see e.g.,][]{Picket96,Toman98}.
Simultaneously, the angular momentum is transferred by a bar mode,
as studied by \citet{Imamura00} and \citet{SHM02}. 
The ambient matter accretes onto this
bar (Fig. 2d), but no ring forms owing to effective
angular momentum transfer. As a result, the bar collapses into
a denser bar (Fig. 2e), eventually reaching 
stellar density 
of $\approx 10^{20}{\rm cm}^{-3}$ (Fig. 2f)
without fragmenting into a multiple system.

To see the dependence on the 
initial perturbation, we ran simulations with $\delta \rho/\rho=0, 10\%$, and 50\%.
As a result, we found that the final results are almost the same regardless of 
the initial perturbations. 
This is because the initial azimuthal perturbation is smoothed out
in the course of the thermal history with $\gamma=1.1$.

\section{Conclusions and Discussion}
\label{Discussion}

We have simulated the collapse of a rotating cylinder while 
assuming the primordial gas thermal history. 
As a result, we have found that to form a binary system,
the formation of a rotationally supported disk is indispensable, which is
realized only by a lower rotation parameter. 
The rotation-supported disk undergoes ring-mode instability,
eventually fragmenting in to binary cores.
In contrast, the case with a relatively high rotation parameter leads to
a bar-mode instability and the angular momentum efficiently transferred by the bar, resulting in the collapse into a single stellar core. 
These results show that the initial conditions with less angular momentum are more prone to binary formation and therefore imply that
a significant fraction of Population III stars are expected to form
in binary systems.

In the present simulation, we have started with high initial density.
If we use the scaling for mass and linear size and
renormalize the physical quantities by a lower initial density 
such as $n_{\rm c0} \la 1 \times 10^{5}{\rm cm}^{-3}$,
then the simulation predicts the formation of a massive binary with several hundred $M_\odot$ and a separation of $\approx 100$AU. 
\citet{NU01} showed that the mass of Population III stars
could bifurcate according to the initial filament density.
If the initial density is above the critical density of H$_2$ molecules,
$n_{\rm c0} > 10^{5}{\rm cm}^{-3}$,
then the fragment mass is peaked around $1M_\odot$,
whereas the peak is shifted to $\approx 100 M_\odot$
for $n_{\rm c0} \la 1 \times 10^{5}{\rm cm}^{-3}$. 

The present simulations show that the binary formation can
occur in the wide mass range of 1 to several $\times 100 M_\odot$, 
if we start with the initial density of 10$^{3}$ -- 10$^{12}$ cm$^{-3}$.

The recent theoretical analyses of the evolution of metal-free 
stars predict that the fate of Population III stars can be classified as follows
 (e.g., Heger \& Woosley 2002): 
\begin{enumerate}
\renewcommand{\labelenumi}{\arabic{enumi}.}
\item A star of $1M_\odot \la M \la 8 M_\odot$ evolves into a white dwarf after asymptotic giant branch mass loss.

\item A star of $8M_\odot \la M \la 25 M_\odot$ results in a Type II SN. 

\
item A star of $25M_\odot \la M \la 140 M_\odot$ probably collapses into a BH.

\item A star of $140M_\odot \la M \la 260 M_\odot$ is partly or completely disrupted by the electron-positron pair instability.

\item A star with mass of $M \ga 260 M_\odot$ collapses completely to a BH without 
ejecting any heavy elements.
\end{enumerate}
Combining such Population III stellar evolution with the present results,
we may expect a wide variety of Population III binaries. 
If a binary is as massive as $\ga 300 M_\odot$, 
it can result in a binary BH. Recently, \citet{Belczynski04} argued 
that a Population III binary BH could be the intensive gravitational wave
burst sources, which may be detected by the Laser Interferometer Gravitational-Wave Observatory. 
If there is a slight mass difference in a massive binary, a system composed
of a $\sim 100 M_\odot$ Population III star and several $100 M_\odot$ BHs 
may form. Then the mass ejected from a massive star can accrete onto 
a companion BH, resulting in a possible source of ionizing radiation
in the early universe \citep{Umemura04}. 
The Population III BH accretion may be an important clue for
the early reionization that is suggested by the {\it Wilkinson Microwave Anisotropy Probe} \citep{RO04a,RO04b}.
If a low-mass Population III binary forms, Population III Type Ia SN might occur. 
In this case, heavy elements synthesized in a more rapidly evolving star 
can pollute a companion star by stellar wind.
The evolution is likely to be sensitive to the 
metallicity-dependent opacity of accretion flow.
Population III Type Ia SNe seem worth studying in the future, because they may play an important role in the thermal history in the universe. 

\acknowledgments
We thank K. Omukai, K. Tomisaka, and T. Tsuribe for stimulating discussion and 
T. Hanawa for contribution to construction of the nested grid code.
The analysis was made with the computational facilities at the National Astronomical Observatory and at the Center for Computational Sciences at the University of Tsukuba.
This research was supported in part by Grants-in-Aid for Specially Promoted Research 16002003 (MU), for Scientific Research (B) 13011204 (MU, TM),
and for Young Scientists (B) 14740134 and 16740115 (TM), 
by the Ministry of Education, Science, Sports, Culture and Technology, Japan



\setcounter{figure}{0}
\clearpage
\begin{figure}
\plotone{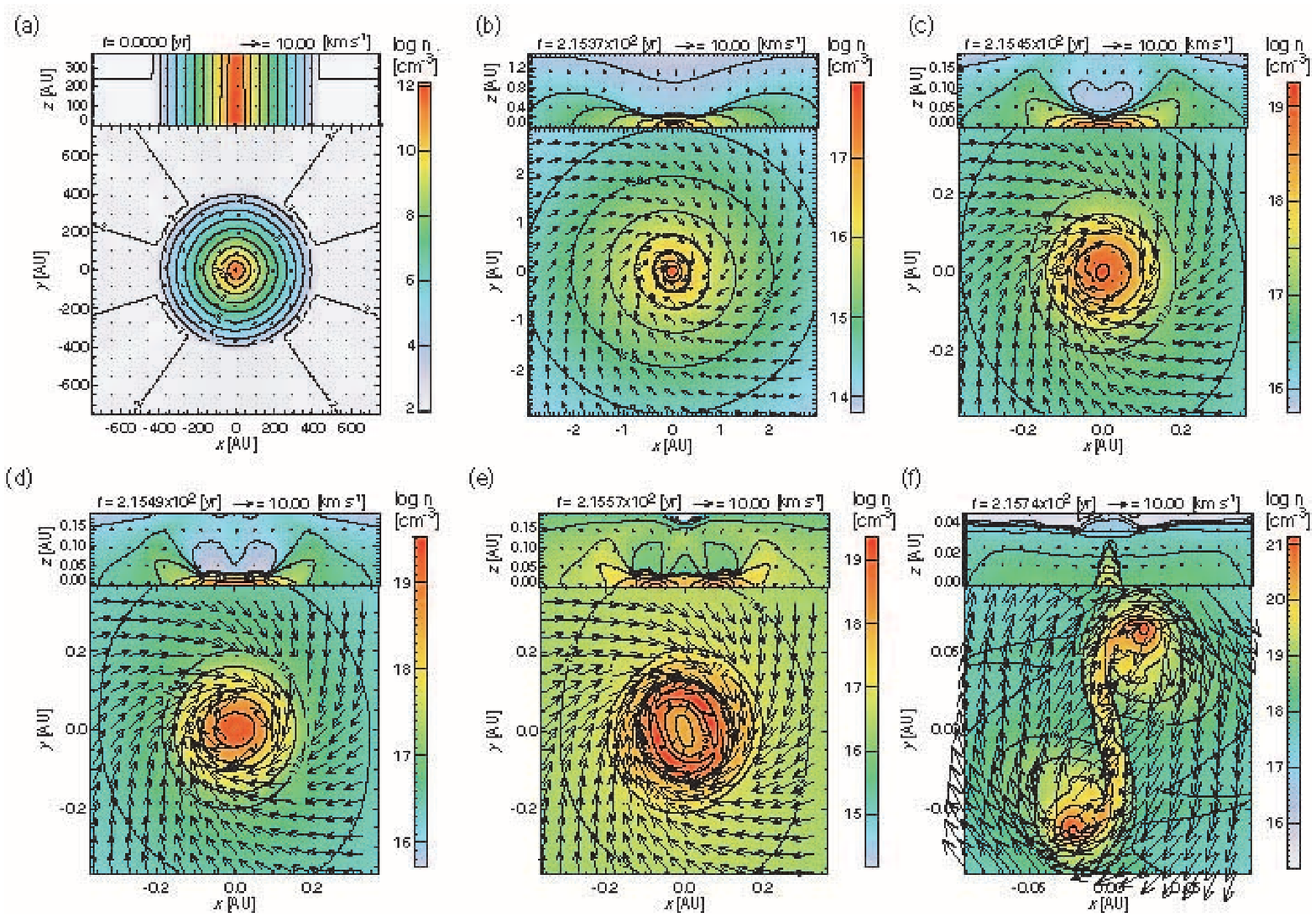}

\caption[dummy]{
Density and velocity distributions in the $x$-$y$ plane on the midplane 
({\it bottom panels}) and $x$-$z$ plane of $y \, = \, 0$ ({\it top panels}) 
for a low $\beta$ model  with $\beta \, = \, 0.1$ at six different stages. 
The color scale and contour curves denote the density in the logarithmic scale. Arrows show the velocity distributions. 
The maximum density and time, ($n_{\rm max}$ [cm$^{-3}$],  $t$ [yr]) is 
({\it a})($1.05 \times 10^{12}$, 0.0), 
({\it b})($8.82 \times 10^{17}$, $2.1537 \times 10^{2}$), 
({\it c})($1.82 \times 10^{19}$, $2.1545 \times 10^{2}$),
({\it d})($6.80 \times 10^{19}$, $2.1549 \times 10^{2}$),
({\it e})($2.35 \times 10^{19}$, $2.1557 \times 10^{2}$), and 
({\it f})($1.32 \times 10^{21}$, $2.1574 \times 10^{2}$)
at each stage.}
\label{fig1}
\end{figure}

\begin{figure}
\plotone{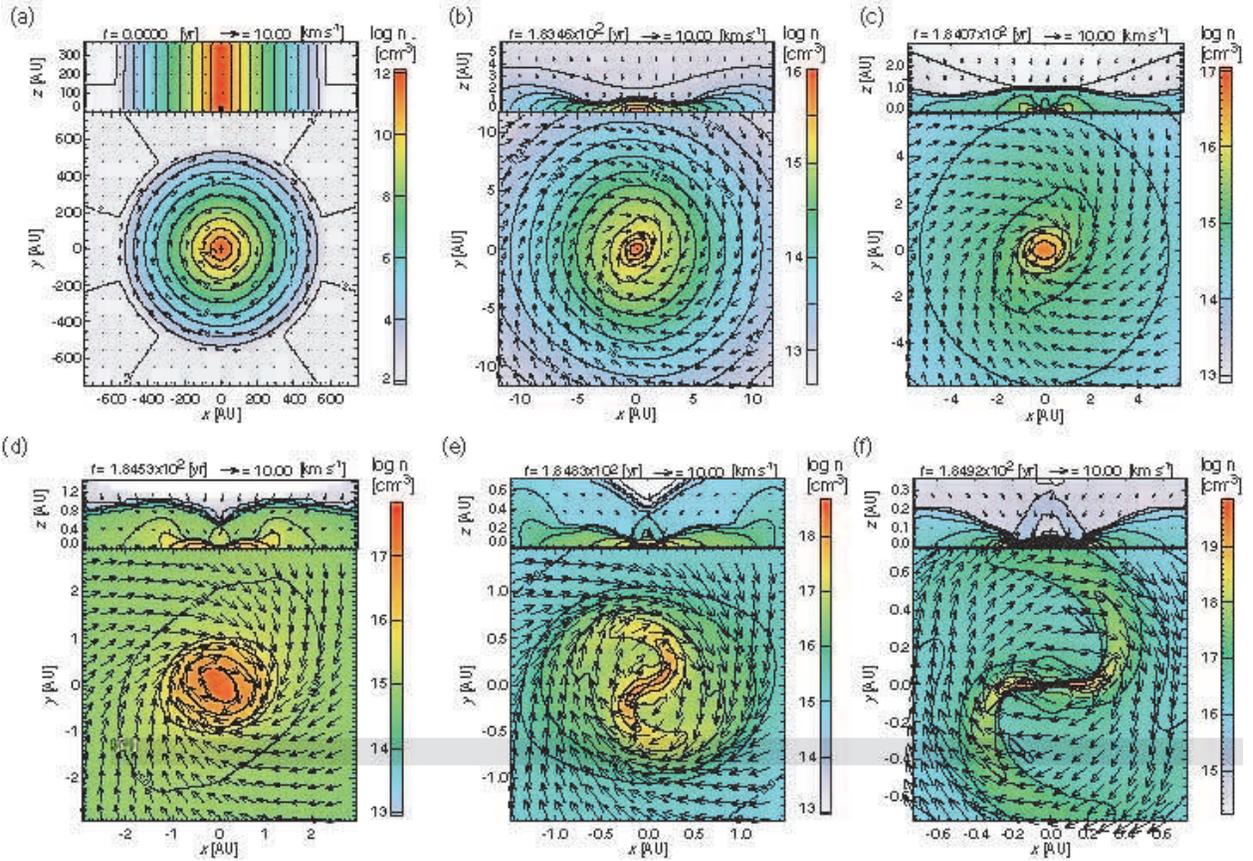}
\caption[dummy]{
Same as Fig. 1 but for a high $\beta$ model with $\beta \, = \, 1.0$.
The maximum density and time, ($n_{\rm max}$ [cm$^{-3}$],  $t$ [yr]) is 
({\it a})($1.05 \times 10^{12}$, 0.0), 
({\it b})($9.93 \times 10^{15}$, $1.8346 \times 10^{2}$), 
({\it c})($1.05 \times 10^{17}$, $1.8407 \times 10^{2}$),
({\it d})($6.89 \times 10^{17}$, $1.8453 \times 10^{2}$),
({\it e})($4.42 \times 10^{18}$, $1.8483 \times 10^{2}$), and 
({\it f})($6.66 \times 10^{19}$, $1.8492 \times 10^{2}$) 
at each stage.
}\label{fig2}
\end{figure}

\end{document}